\documentclass[journal,twocolumn]{IEEEtran}

\usepackage{cite}
\usepackage{amsmath,amsfonts,amsxtra,amssymb,latexsym,amscd,amsthm,mathrsfs,bm}
\newtheorem{lemma}{Lemma}

\usepackage{hyperref}
\usepackage{xcolor}
\usepackage[colorinlistoftodos,textsize=tiny]{todonotes}
\usepackage{soul}
\usepackage{balance}
\usepackage{array}
\usepackage{algorithm}
\usepackage{algpseudocode}
\usepackage{graphicx}
\usepackage{nopageno}
\usepackage{ifpdf}
\ifpdf
\DeclareGraphicsExtensions{.pdf,.png,.jpg}
\else
\DeclareGraphicsExtensions{.eps}
\fi

\usepackage[belowskip=-10pt,aboveskip=5pt,small]{caption}
\IEEEaftertitletext{\vspace{-1.7\baselineskip}}

\makeatletter
\if@todonotes@disabled

\else

\fi
\makeatother

\usepackage{epstopdf}	
\title{\LARGE Deep Learning-based List Sphere Decoding for Faster-than-Nyquist (FTN) Signaling Detection
\thanks{The authors are with the Department of Electrical and Computer Engineering, University of Saskatchewan, Saskatoon, Canada S7N 5A9. Emails: sia942@mail.usask.ca and e.bedeer@usask.ca.}
}

\author{\IEEEauthorblockN{Sina Abbasi and Ebrahim Bedeer}}

\begin{document}
\maketitle

\begin{abstract} Faster-than-Nyquist (FTN) signaling is a candidate non-orthonormal transmission technique to improve the spectral efficiency (SE) of future communication systems. However, such improvements of the SE are at the cost of additional computational complexity to remove the intentionally introduced intersymbol interference. 
In this paper, we investigate the use of deep learning (DL) to reduce the detection complexity of FTN signaling. To eliminate the need of having a noise whitening filter at the receiver, we first present an equivalent FTN signaling model based on using a set of orthonormal basis functions and identify its operation region. Second, we propose a DL-based list sphere decoding (DL-LSD) algorithm that selects and updates the initial radius of the original LSD to guarantee a pre-defined number $N_{\text{L}}$ of lattice points inside the hypersphere. This is achieved by training a neural network to output an approximate initial radius that includes $N_{\text{L}}$ lattice points. At the testing phase, if the hypersphere has more than $N_{\text{L}}$ lattice points, we keep the $N_{\text{L}}$ closest points to the point corresponding to the received FTN signal; however, if the hypersphere has less than $N_{\text{L}}$ points, we increase the approximate initial radius by a value that depends on the standard deviation of the distribution of the output radii from the training phase. Then, the approximate value of the log-likelihood ratio (LLR) is calculated based on the obtained $N_{\text{L}}$ points. Simulation results show that the computational complexity of the proposed DL-LSD is lower than its counterpart of the original LSD by orders of magnitude. 
\end{abstract}

\begin{IEEEkeywords}
Deep learning, Faster-than-Nyquist signaling, list sphere decoding, intersymbol interference, Sequence estimation.
\end{IEEEkeywords}

\IEEEpeerreviewmaketitle

\section{Introduction}
There are increasing demands to improve the spectral efficiency (SE) to meet the requirements of future communication systems. Faster-than-Nyquist (FTN) signaling is a promising candidate technology that can increase the data rate without increasing the transmission bandwidth \cite{ftn}. In  FTN signaling, the data symbols are transmitted at a rate of $1/(\tau T)$, $\tau \leq 1$, when compared to the Nyquist rate of $1/T$ when using $T$-orthogonal pulses, and hence, inter-symbol interference (ISI) is intentionally introduced. 

The early contribution of Mazo \cite{Mazo} showed that increasing the data rate by accelerating the sinc pulses carrying binary phase shift keying (BPSK) symbols up to $\tau = 0.802$ will not deteriorate the asymptotic error rate when compared to Nyquist signaling that operates in the same bandwidth. However, such improvement of the SE is at the cost of prohibitive (at Mazo's time) computational complexity to remove the  introduced ISI. In the past decade, there have been several research works based on conventional signal processing and estimation theory that detect the transmit data symbols of FTN signaling at reduced computational complexity, e.g., \cite{4524864, bedeer2017reduced, bedeer2017very,  kulhandjian2019low}. We refer the reader to \cite{ftn} for a summary of key FTN signaling detection techniques and to \cite{ishihara2021evolution} for a more recent survey. 

Recently, the application of deep learning (DL) to physical layer problems shows promising results mainly when there is a lack of appropriate mathematical models, i.e., model deficit, or a lack of low complexity algorithms, i.e., algorithm deficit \cite{whenML}. Given the fast development of artificial intelligence chips, it is expected that DL will find more applications in physical layers problems. 

The applications of DL have been extended to design FTN signaling systems in \cite{ftnRcv,ftnSP}. In particular, the authors in \cite{ftnRcv} proposed an efficient DL-based architecture for FTN receivers that can replace either the signal detection block or both the signal detection and channel decoding blocks for uncoded and coded  FTN signaling, respectively. Their proposed DL-based FTN receivers showed near optimal performance for non-severe ISI operating regions. In \cite{ftnSP}, the authors proposed a DL-based sum-product algorithm for FTN signaling that operates on a modified factor graph and concatenates a neural network function node to the variable nodes to approximate the optimal error rate performance.

Against the aforementioned literature, in this paper, we investigate the use of DL to reduce the detection complexity of FTN signaling. To eliminate the need of having a noise whitening filter at the receiver, we first present an equivalent transmission model for FTN signaling
with the help of orthonormal basis functions, and we show its operation region. Second, we propose a DL-based list sphere decoding (DL-LSD) algorithm that selects and updates the initial radius of the original LSD to guarantee a pre-defined number $N_{\text{L}}$ of lattice points inside the hypersphere. This is achieved by training a neural network to output an approximate initial radius that includes $N_{\text{L}}$ lattice points. During the testing phase, if the hypersphere has more than $N_{\text{L}}$ lattice points, we keep the $N_{\text{L}}$ closest points to the point corresponding to the received FTN signal; however, if the hypersphere has less than $N_{\text{L}}$ points, we increase the approximate initial radius by a value that depends on the standard deviation of the distribution of the output radii from the training phase. Then, the approximate value of the log-likelihood ratio (LLR) is calculated based on the $N_{\text{L}}$ points. Simulation results show that the average number of flops of the proposed DL-LSD algorithm is three order and one order of magnitude lower than its counterpart of the original LSD, with a selection of the initial radius based on the noise variance, at low and high $E_b/N_0$ values, respectively. 

The rest of the paper is organized as follows. In Section \ref{sec:sys}, we present an equivalent transmission model for FTN signaling {based on using a sum of orthonormal basis}; while in Section \ref{sec:proposed} we discuss the proposed DL-LSD algorithm. Simulation results are presented in Section \ref{sec:sim}, and the paper is concluded in Section \ref{sec:conclusion}.

\begin{figure}[!t]
\centering
\includegraphics[width=0.51\textwidth]{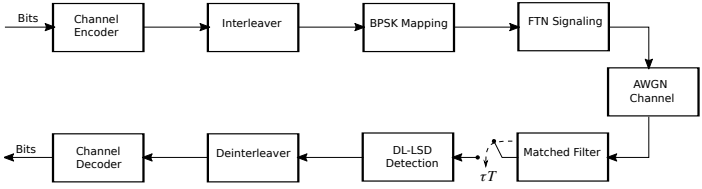}
\vspace*{2pt}
\caption{Block diagram of an FTN signaling system.}
\label{fig:sys}
\end{figure}

\section{System Model and Problem Formulation}\label{sec:sys}
Figure 1 shows a block diagram of an  FTN signaling communication system. At the transmitter side, information bits are encoded, interleaved, and then mapped to data symbols where each symbol is carried by a unit-energy pulse $h(t)$. The widely used FTN signaling model expresses the transmit signal $s(t)$ as:
\begin{IEEEeqnarray}{rCl}
    \label{eq:main1}
    s(t)=\sum_{n} a_n h(t-n\tau T),
\end{IEEEeqnarray}
where $0<\tau \le 1$ is the time acceleration factor, $T$ is the symbol duration, and $a_{n}, n = 1, ..., N,$ is the binary phase shift keying (BPSK) data symbol with average bit energy $E_b$. In our work, we assume that $h(t)$ is a $T$-orthogonal root raised cosine (rRC) pulse with a roll-off factor $\beta_h$. However, such transmission of non-orthogonal pulses in additive white Gaussian noise (AWGN) will require additional discrete-time whitening filter at the receiver to process the colored noise samples after the matched filter. Designing an exact causal and stable discrete-time whitening filter can be challenging at small values of $\tau$ \cite{prlja2008receivers}. To avoid using a whitening filter, one possibility is to use models based on the Ungerboeck observation model that deals directly with the colored noise, e.g. \cite{li2017reduced}. Another possibility which we adopt in this work is to use  an equivalent FTN signaling model that uses a set of orthonormal basis function to  whiten the noise samples after the matched filter. This model appeared in \cite{6241379, textbook} but {has} not received enough attention in the state-of-the-art literature, and it will be discussed here in detail for completeness of the presentation.

In the equivalent FTN signaling model based on orthonormal basis functions, the $T$-orthogonal pulse $h(t)$ is approximated as a sum of $\tau T$-orthonormal pulses $v(t - n \tau T)$ as:
\begin{IEEEeqnarray}{rCl}
    \label{eq:h}
    h(t) \approx \sum_{n} h_n v(t-n\tau T).
\end{IEEEeqnarray} 
In Lemma \ref{lemma:condition}, we discuss how to find the constant coefficient $h_n$ such that the approximation in \eqref{eq:h} is valid.

\begin{lemma}\label{lemma:condition}
For a $T$-orthogonal $h(t)$ pulse, where $H(f) = 0$, $|f|> W$ and $W<0.5/(\tau T)$, let a $\tau T$-orthonormal pulse $v(t)$ have the Fourier transform:
\begin{IEEEeqnarray}{rcl}
V(f) &{}={}& \left\{\begin{matrix}
C_o, &  |f|<W,\\
0, &  |f|>\frac{1}{\tau T}-W,\\ 
\end{matrix}\right.
\end{IEEEeqnarray}
where $C_o$ is a constant. Then $h(t)$ may be expressed as $h(t) = \sum_{n} h_n v(t-n\tau T),$ where:
\begin{IEEEeqnarray}{rCl}
\label{eq:lemmah}
h_{n} =\frac{\tau T}{C_o} h(n\tau T).
\end{IEEEeqnarray}
\end{lemma}
\emph{Proof:} see Appendix.

\begin{figure}[!t]
\centering
\includegraphics[width=0.51\textwidth]{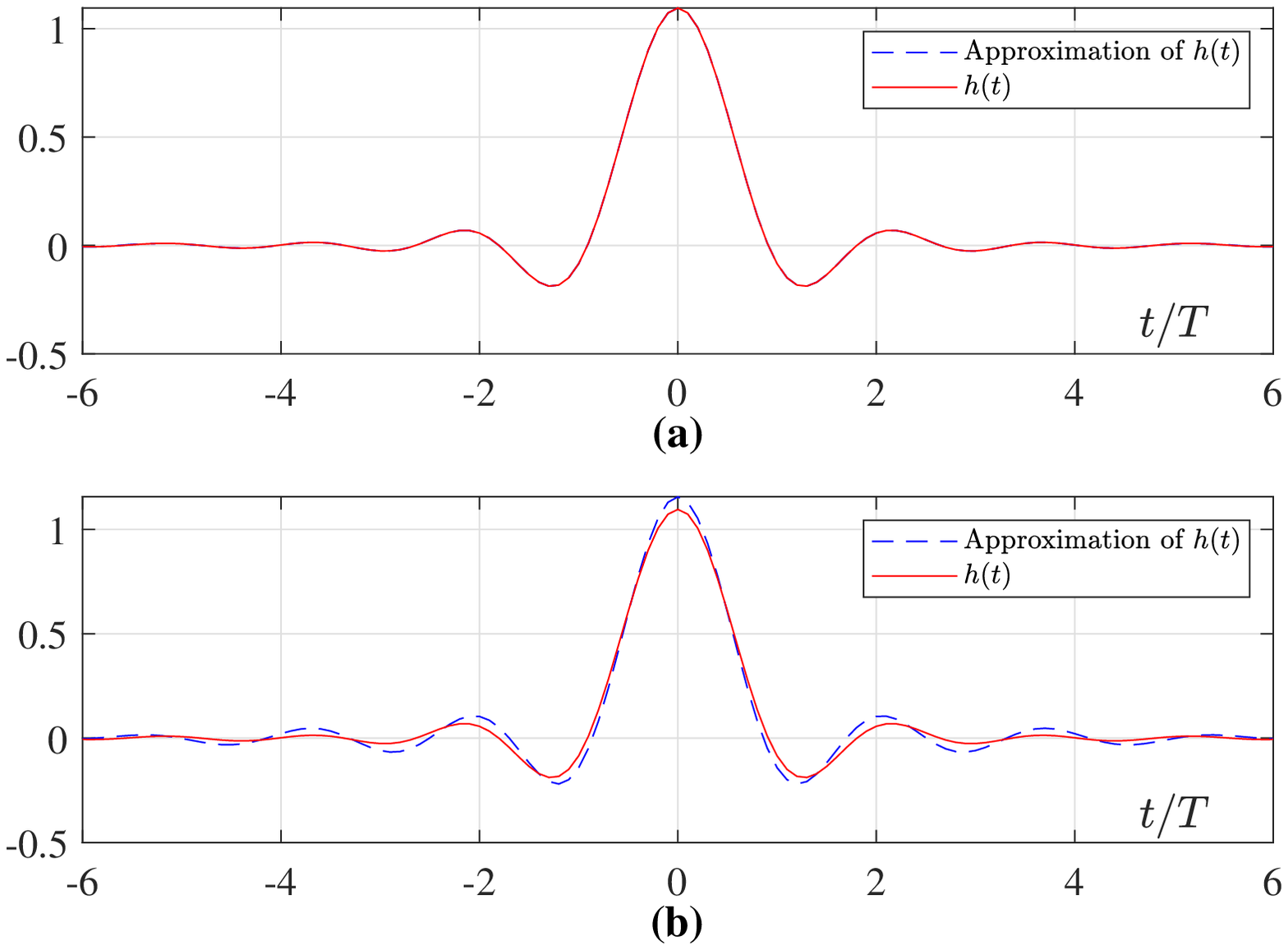}
\caption{(a) $\tau = 0.6$, (b) $\tau = 0.9$. The solid line is the exact $h(t)$ pulse and the dashed line is its approximation based on (\ref{eq:h}).}
\label{fig:tau}
\end{figure}

As one can see from Lemma \ref{lemma:condition},  $h(t)$ can be approximated as a sum of $\tau T$-orthogonal basis functions $v(t - n \tau T)$ weighted by the scaled samples of $h(t)$  in \eqref{eq:lemmah}, if $W<0.5/(\tau T)$ and $V(f)$ is constant for $|f|<W$. For example and as shown in Fig. \ref{fig:tau} (a), when $h(t)$ is a $T$-orthonormal rRC with a roll-off factor $\beta_h = 0.35$ with a bandwidth $W=0.5(1+\beta_{h})/T$, it can be represented as a sum of 20 rRC $\tau T$-orthonormal pulses, i.e., $\sum_{n=1}^{20} h_n v(t-n\tau T)$, with a roll-off factor  $\beta_v = 0.12$ and $\tau = 0.6$, if $W < 0.5/(\tau T)$, which yields: 
\begin{IEEEeqnarray}{rCl}
\label{eq:cons1}
\tau < \frac{1}{1+\beta_{h}}.
\end{IEEEeqnarray}
Hence, the condition in \eqref{eq:cons1} defines the operation region of the FTN signaling equivalent model. 
On the other hand, in Fig. \ref{fig:tau} (b), $\tau = 0.9$ does not satisfy \eqref{eq:cons1}, and hence, the approximation is not accurate.

Given Lemma \ref{lemma:condition} and substituting \eqref{eq:h} in \eqref{eq:main1}, the equivalent FTN signaling transmit signal using the orthonormal basis function is expressed as:
\begin{IEEEeqnarray}{rCl}
\label{eq:newMain1}
s(t) = \sum_{n} b_n v(t-n\tau T),
\end{IEEEeqnarray}
where $b_n = \sum_{l} a_{n-l}  h_l$
and $h_{l}$ is given in (\ref{eq:lemmah}).
Assuming AWGN channel, the received signal is passed through a filter matched to the orthonormal basis $v(t)$, 
and for a real and symmetric $v(t)$, it is given as: 
\begin{IEEEeqnarray}{rCl}
y(t) = (s(t)+w(t)) \ast v(t),
\end{IEEEeqnarray}
where $w(t)$ is additive white Gaussian noise (AWGN) with zero mean and variance of $\sigma^{2}$ and $\ast$ denotes the convolution. Then, this signal is sampled every $\tau T$ and is written as:
\begin{IEEEeqnarray}{rCl}
    \label{eq:rec1}
     y_{n} = b_{n} + w_{n}.
\end{IEEEeqnarray}
The received sampled FTN signal can be expressed in a matrix form as:
\begin{IEEEeqnarray}{rCl}
    \label{eq:rec2}
     \bm{y =  Ha + w},
\end{IEEEeqnarray}
where $\bm a$ and $\bm w$ are the transmit data symbol and white noise vectors, respectively.  

The received vector $\bm y$ needs to be processed by an FTN signaling detector to produce a soft output that can be used by the channel decoder. 
This soft output can be obtained from maximizing a posteriori probability (APP) for a given bit, and it is expressed as a log-likelihood ratio (LLR) value. 
The LLR for a bit $x_{k}$ given the received vector $\bm{y}$ is written as:
\begin{IEEEeqnarray}{rCl}
L_{D}\left(x_{k} \mid \bm{y}\right)=\log \frac{P\left(\operatorname{map}(x_{j})=+1 \mid \bm{y}\right)}{P\left(\operatorname{map}(x_{j})=-1 \mid \bm{y}\right)},
\label{eq:LLR1}
\end{IEEEeqnarray}
where $x_{k}$ is the $k$th bit of $N \times 1$ vector $\bm{x}$ of all bits in one transmit block. We map the binary bits of 0 and 1 to $-1$ and $+1$, respectively. 
Assuming that $x_{k}, k=0, ..., N-1$, are statistically independent, we use the Bayes theorem to re-write  (\ref{eq:LLR1}) as \cite{LLR}:
\begin{IEEEeqnarray}{lLl}
\label{eq:LLR2}
L_{D}\left(x_{k} \mid \bm{y}\right) = \\ \nonumber
L_{A}\left(x_{k}\right)+\ln \frac{\sum_{\bm{x} \in \bm{\mathcal{X}}_{k},+1} p(\bm{y} \mid \bm{x}) \cdot \exp \sum_{j \in \mathcal{J}_{k, \bm{x}}} L_{A}\left(x_{j}\right)}{\sum_{\bm{x} \in \bm{\mathcal{X}}_{k,-1}} p(\bm{y} \mid \bm{x}) \cdot \exp \sum_{j \in \mathcal{J}_{k, \bm{x}}} L_{A}\left(x_{j}\right)},
\end{IEEEeqnarray}
where {$\bm{\mathcal{X}}$ is the set of all $2^N$ possible lattice points $\bm{x}$}, $\bm{\mathcal{X}}_{k,+1} = \{\bm{x} \mid x_{k} =+1\}$, $\bm{\mathcal{X}}_{k,-1} = \{\bm{x} \mid x_{k} =-1\}$, $\mathcal{J}_{k, \bm{x}}=\{j|j=0,..., N-1, j\ne k, x_{j} = 1\}$, and
\begin{IEEEeqnarray}{lLl}
\label{eq:LLR3}
L_{A}(x_{j})=\ln \frac{P(\operatorname{map}(x_{j})=+1)}{P(\operatorname{map}(x_{j})=-1)},
\end{IEEEeqnarray}
and the likelihood function $p(\bm{y} \mid \bm{x})$ is given as follow:
\begin{IEEEeqnarray}{lLl}
\label{eq:LLR5}
p(\bm{y} \mid \bm{x})=\frac{\exp \left(-\frac{1}{2 \sigma^{2}} \cdot\|\bm{y}-\bm{H}\bm{a}\|^{2}\right)}{\left(2 \pi \sigma^{2}\right)^{N}}.
\end{IEEEeqnarray}

\section{Proposed DL-LSD Algorithm}\label{sec:proposed}

\subsection{Review of The LSD Algorithm}
Calculation of the LLR value for each bit in (\ref{eq:LLR2}) needs to consider the whole possible lattice points in $\bm{\mathcal{X}}$, which has the size of $2^{N}$ of the skewed lattice points. Since for each bit $x_k$ we iterate over all lattice points in $\bm{\mathcal{X}}$ and the calculation inside the $\exp$ function takes $O(N)$, and each transmit block has $N$ bits in total; then, the computational complexity of the LLR values of one transmit block is at the order of $O(2^{N} N^2)$. For example, when the transmission block has $N=25$ symbols; then the set $\bm{\mathcal{X}}$ has $2^{25}$ $N$-dimensional points. Accordingly, calculating the (\ref{eq:LLR2}) for all bits within the transmit block requires $25 \times 2^{25} \approx 2\times10^{10}$ operations.

One can see from (\ref{eq:LLR5}) that the conditional probability $p(\bm{y}\mid\bm{x})$ has an exponential relation with the distance of the skew lattice points to $\bm{y}$, i.e., $\|\bm{y}-\bm{H}\bm{a}\|^{2}$. That said, to reduce the complexity of the calculations of the LLR values in (\ref{eq:LLR2}), we can consider a pre-defined number of points close to $\bm{y}$ rather all possible points in $\bm{\mathcal{X}}$. Finding the closest number of pre-defined points to $\bm{y}$ can be obtained by modifying the SD to what is called the LSD \cite{LLR}. The LSD finds the first $N_{\text{L}}$ closest lattice points in the skew lattice $\bm{Ha}$ to the vector $\bm{y}$ corresponding to the received FTN signaling, and then, forms the candidate list $\bm{\mathcal{L}}$. 

To form the candidate list $\bm{\mathcal{L}}$, the SD is modified as follows. When a lattice point is found inside the hypersphere, the initial radius of the hypersphere is not reduced to the distance of that lattice point; rather, we add this lattice point to our list $\bm{\mathcal{L}}$. However, if the size of $\bm{\mathcal{L}}$ became $N_{\text{L}} + 1$, the lattice point with the largest distance to the vector $\bm{y}$ in $\bm{\mathcal{L}}$ is removed and the radius  is updated to the largest distance to the vector $\bm{y}$ among all the remaining $N_{\text{L}}$ lattice points in $\bm{\mathcal{L}}$. At the end and instead of using all the lattice points in $\bm{\mathcal{X}}$, the LSD algorithm finds the $N_{\text{L}}$ closest points to the vector $\bm{y}$ that are to be used in the calculations of the approximate LLR values as follows: 
\begin{IEEEeqnarray}{lLl}
\label{eq:LLRapprox}
\tilde{L}_{D} \left(x_{k} \mid \bm{y}\right) = \\ \nonumber
\tilde{L}_{A}\left(x_{k}\right)+\ln \frac{\sum_{\bm{x} \in \bm{\mathcal{L}}_{k},+1} p(\bm{y} \mid \bm{x}) \cdot \exp \sum_{j \in \mathcal{J}_{k, \bm{x}}} \tilde{L}_{A}\left(x_{j}\right)}{\sum_{\bm{x} \in \bm{\mathcal{L}}_{k,-1}} p(\bm{y} \mid \bm{x}) \cdot \exp \sum_{j \in \mathcal{J}_{k, \bm{x}}} \tilde{L}_{A}\left(x_{j}\right)},
\end{IEEEeqnarray}
where $\bm{\mathcal{L}}_{k,+1} = \{\bm{x} \in \bm{\mathcal{L}}  \mid x_{k} =+1\}$, $\bm{\mathcal{L}}_{k,-1} = \{\bm{x} \in \bm{\mathcal{L}} \mid x_{k} =-1\}$. Also, $\tilde{L}_{A}$ is obtained similar to $L_{A}$ but by considering lattice points inside $\bm{\mathcal{L}}$ instead of the whole lattice $\bm{\mathcal{X}}$.

As can be seen from \eqref{eq:LLRapprox}, the computational complexity to approximate the LLR value of each transmit block of symbols reduces to $O(N_L N^2)$ because we consider the $N_{\text{L}}$ elements in $\bm{\mathcal{L}}$ instead of whole $2^{N}$ lattice points. For example, if we consider $N_L=32$ and $N=25$, the calculation of (\ref{eq:LLRapprox}) for all bits within the transmit block requires $25\cdot25^2\approx1.5\times10^{3}$ operations which is way less than $2\times10^{10}$ required to calculate the exact LLRs.

On one hand, selecting the initial radius of the LSD to be of large value will lead to a comparable complexity to the exhaustive search due to the large number of lattice points inside the hypersphere. On the other hand, selecting the initial radius to be of small value may not guarantee to have $N_{\text{L}}$ lattice points, and hence, degrade the approximation quality of the LLR values in \eqref{eq:LLRapprox}. Hence, it is clear from the previous discussion that the selection of the initial radius of the LSD to have $N_{\text{L}}$ lattice points is crucial to reduce its tree search complexity while maintaining an acceptable approximation of the LLR values. That said, we propose a DL-LSD algorithm to find the proper initial radius that guarantees to have $N_{\text{L}}$ lattice points.

\subsection{The Training Phase of the Proposed DL-LSD Algorithm}
The intuition behind our proposed DL-LSD algorithm is estimating the initial radius to guarantee a pre-defined number $N_{\text{L}}$ of lattice points inside the hypersphere. Similar idea for estimating the initial radius that guarantees at least one point inside the hypersphere has been proposed in \cite{mostafa}.  
This radius estimation problem is a non-linear regression problem, and neural networks (NNs) have shown success in solving such problems \cite{DLBook}. That said, we propose to use NNs to predict the initial radius that guarantees to include a pre-defined number $N_{\text{L}}$ of lattice points to the received FTN signaling vector $\bm{y}$.The training data are obtained from the implementation of the LSD. Then, we feed the NN with the received vector $\bm{y}$ as an input, and we consider the distance of the furthest point in $\bm{\mathcal{L}}$ from $\bm{y}$ as the desired radius for training the output of the NN. Therefore, the set of input-output pairs $\{\bm{y}^{(i)},R^{(i)}\}$ is used to train our NN, where $R^{(i)}$ is the largest radius in $\bm{\mathcal{L}}^{(i)}$, and $i = 1, ..., S$, where $S$ is the size of training data set. The NN, $f$, predicts the initial radius $\hat{R}$ at its output layer as:
\begin{IEEEeqnarray}{rCl}
\label{eq:NN1}
    \hat{R}^{(i)}=f(\bm{y}^{(i)},\theta),
\end{IEEEeqnarray}
where $\theta$ is the set of all parameters of NN, i.e, weights and biases values. {Please note that the input to the NN $\bm{y}^{(i)}$ captures the effect of the ISI in $\bm{H}$ based on \ref{eq:rec2}. Since we train the NN for each value of $\tau$, for which the ISI matrix $\bm{H}$ will be the same for all training data, we decided to not feed the NN with $\bm{H}$ directly.}

The first and last layers are the input and output layers, respectively; while the three middle layers are the hidden layers. The first two hidden layers are recurrent neural network (RNN) layers with 128 neurons and a simple fully connected layer with 64 neurons is used as the third hidden layer. Please note that the number of hidden layers and the number of neurons in each layer has been chosen experimentally. We use the activation function $\text{Relu}$  for all hidden layers and it is defined as $\text{Relu}(u)=\max(0,u)$.
We use the mean square error (MSE)  to evaluate the prediction error of the initial radius, and it is defined as:
\begin{IEEEeqnarray}{rCl}
\label{eq:NN3}
    L(\theta)=\frac{1}{\left|S\right|} \sum_{i=1}^{S}\left(R^{(i)}-f\left(\bm{y}^{(i)}, \theta\right)\right)^{2},
\end{IEEEeqnarray}
where the desired radius $R^{(i)}$ is output when $\bm{y^{(i)}}$ is used as an input. An approximation of (\ref{eq:NN3}) in each iteration $t$ over one training epoch can obtain as follow:
\begin{IEEEeqnarray}{rCl}
\label{eq:NN4}
    \tilde{L}_{t}(\theta)=\frac{1}{\left|S_{t}\right|} \sum_{i \in S_{t}}\left(R^{(i)}-f\left(\bm{y}^{(i)}, \theta\right)\right)^{2},
\end{IEEEeqnarray}
where we divide our data set $S$ to $B$ mini-batches, each mini-batch $S_{t}$ has a size of $|S_{t}|=|S|/B$. The complexity of the gradient computation is remarkably reduced when we increase the number of mini-batches $B$, while the variance of updating the NN parameter, i.e., $\theta$, still decreases.
Finally, Adam \cite{adam} is used as an optimization method for updating  $\theta$.

\subsection{The Testing Phase of the Proposed DL-LSD Algorithm}

\begin{figure}[!t]
\centering
\includegraphics[width=0.50\textwidth]{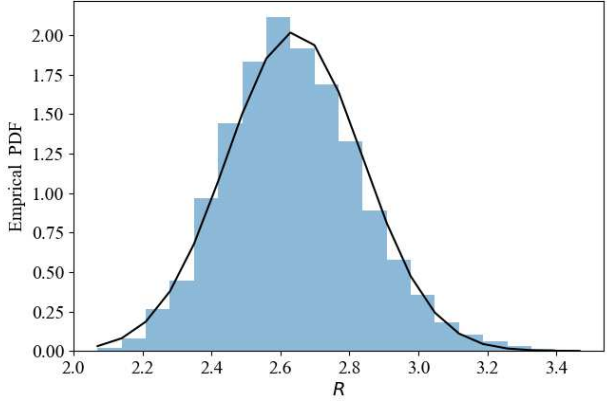}
\caption{The histogram of the obtained radii of the training phase for $\tau=0.6$, $\beta_h=0.35$, $\beta_v=0.12$, and $E_{b}/N_{0}=8$ dB.}
\label{fig:his}
\end{figure}

In the testing phase, the received FTN signaling $\bm{y}$ is fed to the trained NN, and the NN gives the estimation of initial radius $\hat{R}$ that approximately guarantees to have $N_{\text{L}}$ lattice points inside the hypersphere centered at $\bm{y}$. Then the LSD algorithm is executed  with an initial radius equal to the obtained initial radius from the NN, i.e.,  $d=\hat{R}$. However, there is a possibility that the $d$ is large enough to include more than $N_{\text{L}}$ points at the testing phase. In this case, we keep only the $N_{\text{L}}$ points with the smallest radii from $\bm{y}$ and discard the extra points with the largest radii. In case  $d$ is not large enough to have at least $N_{\text{L}}$ points at the testing phase, we propose to increase the radius $d$ by a value $\delta_d$, and then, execute the LSD algorithm with new radius $d + \delta_d$. The selection of $\delta_d$ can be explained with the help of Fig. \ref{fig:his} as follows. In Fig. \ref{fig:his}, we sketch the empirical probability density function (PDF) of all the obtained radii from the training phase at $\tau = 0.6$, $\beta_h=0.35$, $\beta_v=0.12$, and $E_b/N_o = 8$. We observe that the empirical PDF can be approximated as a Gaussian distribution with a standard deviation $\delta_d$. That said, in case the initial radius $d$ has less than $N_{\text{L}}$, we increase the radius by $\delta_d$. The proposed DL-LSD algorithm is summarized at the top of this column. Finally, the approximate LLR values are calculated according to \eqref{eq:LLRapprox}, and then passed to the channel decoder as soft inputs to estimate the transmit data symbols $\bm{\hat{a}}$.

\begin{algorithm}[!t]
\caption*{The Proposed DL-LSD Algorithm}\label{alg:algo}
\begin{algorithmic}
\State $\textbf{Input:} \: \bm{H},\:\bm{y},\:\delta_d,\: f(.\:,\bm{\theta})$
\State $\textbf{Output:}\: $ Calculated LLR values
\State $d \leftarrow f(\bm{y},\bm{\theta})$ \Comment{Estimating radius with NN}
\While {True}
\State $\bm{\mathcal{L}} \leftarrow \operatorname{LSD}(\bm{H},\:\bm{y},\:d)$ \Comment{LSD algorithm returns a list}
\If {$|\bm{\mathcal{L}}| < N_{\text{L}}$}
    \State $d = d + \delta_d$ \Comment{Increasing radius}
\Else
    \State $\bm{\mathcal{L}} = \bm{\mathcal{L}}(1:N_\text{L})$ \Comment{Picking first $N_\text{L}$ closet point to $\bm{y}$}
    \State \textbf{break} \Comment{breaking the while loop}
\EndIf
\EndWhile
\State $\operatorname{LLR}(\bm{\mathcal{L}})$ \Comment{Calculation of LLR based on (\ref{eq:LLRapprox})}

\end{algorithmic}
\end{algorithm}

\section{Simulation results}\label{sec:sim}
In this section, we investigate the performance of the proposed DL-LSD to detect coded BPSK FTN signaling. We consider a standard convolutional code (7, [171 133]) to encode the information bits at the transmitter and a Viterbi decoder to decode the approximate soft outputs of the proposed DL-LSD at the receiver. The roll-off factors $\beta_h$ and $\beta_v$ are set to 0.35 and 0.12, respectively. We consider $N=25$ data symbols per block transmission and an acceleration factor of $\tau$ = 0.6 and 0.74. Please note that the choice of these values of $\tau$ meets the condition in \eqref{eq:cons1}.

The training of the proposed DL-LSD can be summarized as follows. For $E_{b}/N_{0}=$ 4 and 6 dB, we use 800 blocks of random data symbols; while for $E_{b}/N_{0}=$ 8 and 10 dB, we use 8000 blocks. For each of the training blocks, the number of random data symbols per block is set to $N=25$. Please note that the low number of blocks used to train the NN at $E_{b}/N_{0}=$ 4 and 6 dB is due to the huge computational complexity required to obtain the training data symbols from the original-LSD that selects the initial radius based on the noise variance \cite{hassibi1}. We experimentally set the learning rate of Adam optimizer to 0.0001,  and the mini-batch size $S_t$ is set to 20.

\begin{figure}[!t]
\centering
\includegraphics[width=0.51\textwidth]{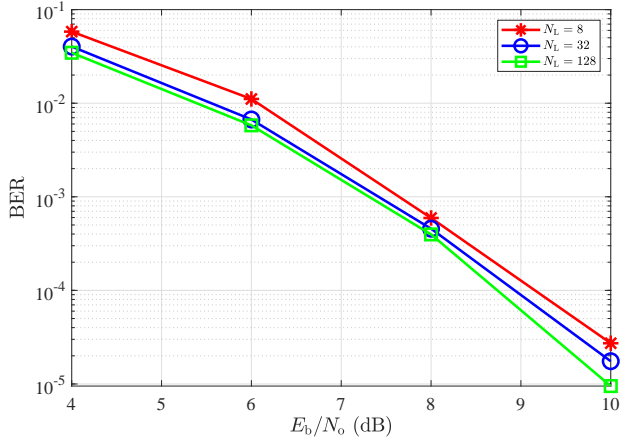}
\caption{BER as a function of $E_{b}/N_{0}$ at $\tau=0.6$ for different values of $N_{\text{L}}$.}
\label{fig:LLR}
\end{figure}

As discussed earlier, the aim of the proposed DL-LSD algorithm is to select a number of lattice points $N_{\text{L}}$ to approximate the calculations of the LLR value of each bit in \eqref{eq:LLRapprox} without deteriorating the error rate performance. Comparisons of the performance of the sphere decoding with other low complexity FTN signaling detection techniques can be found in \cite{bedeer2017reduced, ibrahim2021novel}.
To strike a balance between the computational complexity and the BER performance, we plot in Fig. \ref{fig:LLR} the BER performance for different values of $N_{\text{L}}$ at $\tau =0.6$. As can be seen, the value of $N_L=32$ shows negligible BER loss when compared to $N_L=128$, while significantly reduces the complexity of calculating \eqref{eq:LLRapprox}. Hence, we adopt the value of $N_{\text{L}}$ to be 32 in the rest of the simulation results.

\begin{figure}[!t]
\centering
\includegraphics[width=0.51\textwidth]{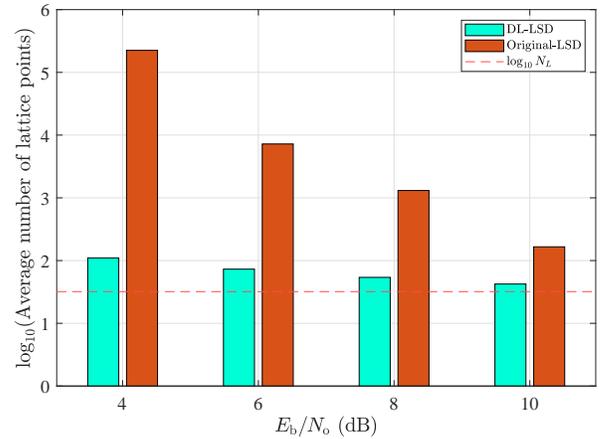}
\caption{Comparison of the average number of lattice points inside the hypersphere of the proposed DL-LSD algorithm and the original-LSD at $\tau=0.6$.}
\label{fig:lNum}
\end{figure}

Fig. \ref{fig:lNum} depicts the average number of lattice points inside the hypersphere of both the original-LSD and proposed DL-LSD versus $E_b/N_0$ for $\tau = 0.6$. Please note that the average number of lattice points is calculated based on averaging the results of 10 transmit blocks. As can be seen, the average number of lattice points obtained by the proposed DL-LSD algorithm is close to the target value of $N_L = 32$, and more importantly, is insensitive to the noise power. This is in contrast to the original-LSD where the initial radius is set based on the noise variance \cite{hassibi1}, and hence, can have a large number of lattice points inside hypersphere at low $E_b/N_0$. Since the complexity of the tree search exponentially increases with increasing the number of lattice points inside the hypersphere, the proposed DL-LSD is expected to have a reduced complexity when compared to the original-LSD. 

\begin{figure}[!t]
\centering
\includegraphics[width=0.51\textwidth]{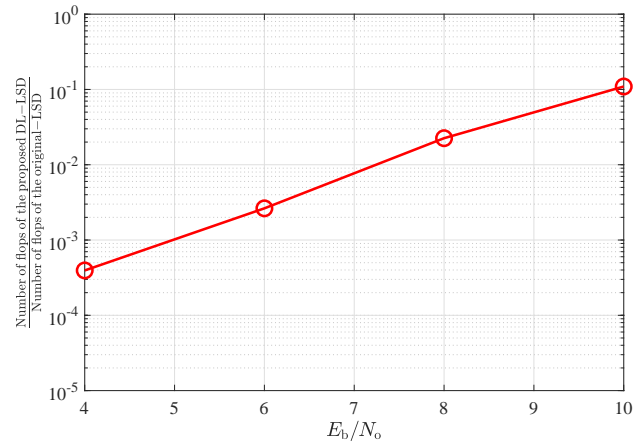}
\caption{Comparison of the average number of flops of the proposed DL-LSD algorithm and the original-LSD at $\tau=0.6$.}
\label{fig:flops}
\end{figure}

To quantify the reduction of the computational complexity of the proposed DL-LSD algorithm with respect to the original-LSD, in Fig. \ref{fig:flops} we plot the ratio of the number of floating point operations (flops) of the proposed DL-LSD to the number of flops in original-LSD as a function of $E_b/N_0$ for $\tau = 0.6$. A flop serves as a basic unit of computation, and it denotes one addition, subtraction, multiplication, or division of floating point numbers. To have a fair complexity comparison, both DL-LSD and the original-LSD use the same implementation of the LSD algorithm but they are different only in the selection of the initial radius (the proposed DL-LSD algorithm estimates the initial radius from the trained NN, while the original-LSD estimates the initial radius based on the noise variance as in \cite{hassibi1}). As one can see, the proposed DL-LSD algorithm has more than three orders of magnitude lower number of flops when compared to the original-LSD algorithm for low values of $E_b/N_0$. For high values of $E_b/N_0$, the proposed algorithm achieves an order of magnitude lower number of flops.

\begin{figure}[!t]
\centering
\includegraphics[width=0.51\textwidth]{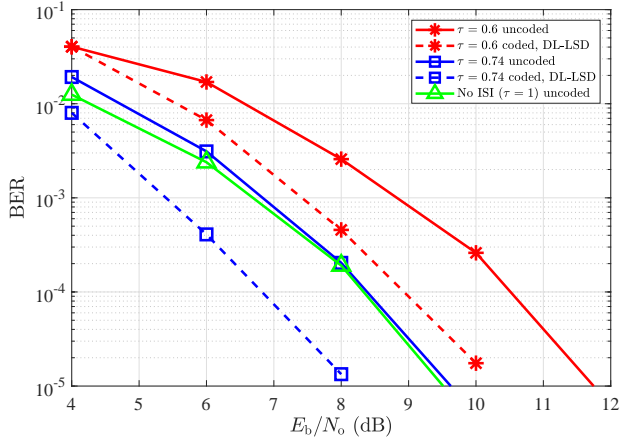}
\caption{Coded and uncoded BER performance for different values of $\tau$.}
\label{fig:Ber}
\end{figure}

In Fig. \ref{fig:Ber}, we depict the BER of the uncoded and coded FTN signaling as a function of $E_{b}/N_{0}$ for different values of $\tau$. 
As one can see, for the uncoded transmission at $\tau = 0.74$, the BER approaches its counterpart of Nyquist signaling which represents 35\% in the SE at no increase in $E_b/N_o$. Decreasing the value of $\tau$ will results in an improvement in the SE but at the cost of increasing $E_b/N_o$. For the coded transmission, the proposed DL-LSD showed approximately savings of 1.5 dB in $E_b/N_o$ when compared to the uncoded results at both $\tau = 0.74$ and $0.6$ at BER of $10^{-4}$.

\section{Conclusion}\label{sec:conclusion}
FTN signaling can improve the SE without increasing the transmission bandwidth, and hence, it is a promising technology for future communication systems. In this paper, we presented an equivalent transmission model for FTN signaling that uses a set of orthonormal basis functions to eliminate the need to design a noise whitening filter at the receiver. 
We then proposed a DL-LSD algorithm to learn and update an approximate initial radius to include a certain number of points $N_{\text{L}}$ inside the hypersphere. In case the initial radius has less than $N_{\text{L}}$ points, we increase the approximate initial radius by a value that depends on the standard deviation of the distribution of the output radii from the training phase. Simulation results showed that the average number of flops of the proposed DL-LSD algorithm is three order and one order of magnitude lower than its counterpart of the original LSD, with a selection of the initial radius based on the noise variance, at low and high $E_b/N_0$ values, respectively.

\section*{Appendix\\Proof of Lemma 1}
 
The proof appears in \cite{textbook} and it is included here for completeness of the presentation.
We define the discrete-time Fourier transform $H_{s}(f_d) = \sum_n h_n e^{-j2\pi f_d n}$, where $\{h_n\}$ is the sampled sequence of $h(t)$ every $\tau T$. From the properties of the Fourier transform we know that $H_s(f) = \frac{1}{\tau T}\sum_{n} H(f- \frac{n}{\tau T})$, where $H(f)$ is the continuous-time Fourier transform of $h(t)$, hence: 
\begin{IEEEeqnarray}{rCl}
    \label{eq:lemma2}
    H(f) &=& \tau T H_{s}(f) 
    \nonumber\\
    &=& \tau T \sum_{n} h(n\tau T)e^{-j2\pi fn\tau T},\quad |f| \le \frac{1}{\tau T} - W. \IEEEeqnarraynumspace
\end{IEEEeqnarray}
At the same time, from the definition of inverse continues-time Fourier transform  we have $h(t) \triangleq \int H(f) e^{j2\pi ft} df$.
Considering the fact that $V(f)$ is constant over the support of $H(f)$, i.e., $|f|<W$. Then, we can multiply $V(f)$ inside the integral and divide by the constant $C_{0}$ outside of the integral to have:
\begin{IEEEeqnarray}{rCl}
    \label{eq:lemma3}
    h(t) = \frac{1}{C_{0}} \int H(f) V(f) e^{j2\pi ft} df.
\end{IEEEeqnarray}
Since $V(f)=0$ for $|f| \ge \frac{1}{\tau T} - W$, substituting  (\ref{eq:lemma2}) into (\ref{eq:lemma3}) results in:
\begin{IEEEeqnarray}{rCl}
    \label{eq:lemma4}
    h(t) &=&\frac{1}{C_{0}} \int \left[ \tau T \sum_{n} h(n\tau T)e^{-j2\pi fn\tau T} \right] V(f) e^{j2\pi ft} df
    \nonumber\\
    & = & \sum_{n} \left[ \frac{\tau Th(n\tau T)}{C_{0}} \right]\int V(f) e^{j2\pi f(t-n\tau T)} df \nonumber \\
    &=& \sum_{n} h_{n}v(t-n\tau T),
\end{IEEEeqnarray}
where $h_{n} = \frac{\tau T}{C_{0}} h(n\tau T)$ which concludes the proof. \hfill$\blacksquare$

\bibliographystyle{IEEEtran} 
\bibliography{IEEEabrv,refs} 

\begin{thebibliography}{10}
\providecommand{\url}[1]{#1}
\csname url@samestyle\endcsname
\providecommand{\newblock}{\relax}
\providecommand{\bibinfo}[2]{#2}
\providecommand{\BIBentrySTDinterwordspacing}{\spaceskip=0pt\relax}
\providecommand{\BIBentryALTinterwordstretchfactor}{4}
\providecommand{\BIBentryALTinterwordspacing}{\spaceskip=\fontdimen2\font plus
\BIBentryALTinterwordstretchfactor\fontdimen3\font minus
  \fontdimen4\font\relax}
\providecommand{\BIBforeignlanguage}[2]{{%
\expandafter\ifx\csname l@#1\endcsname\relax
\typeout{** WARNING: IEEEtran.bst: No hyphenation pattern has been}%
\typeout{** loaded for the language `#1'. Using the pattern for}%
\typeout{** the default language instead.}%
\else
\language=\csname l@#1\endcsname
\fi
#2}}
\providecommand{\BIBdecl}{\relax}
\BIBdecl

\bibitem{ftn}
J.~B. Anderson, F.~Rusek, and V.~Öwall, ``Faster-than-{N}yquist signaling,''
  \emph{Proc. IEEE}, vol. 101, no.~8, pp. 1817--1830, Mar. 2013.

\bibitem{Mazo}
J.~E. Mazo, ``Faster-than-{N}yquist signaling,'' \emph{The Bell System
  Technical Journal}, vol.~54, no.~8, pp. 1451--1462, Oct. 1975.

\bibitem{4524864}
F.~{Rusek} and J.~B. {Anderson}, ``Non binary and precoded faster than
  {N}yquist signaling,'' \emph{IEEE Trans. Commun.}, vol.~56, no.~5, pp.
  808--817, May 2008.

\bibitem{bedeer2017reduced}
E.~Bedeer, H.~Yanikomeroglu, and M.~H. Ahmed, ``Reduced complexity optimal
  detection of binary faster-than-{N}yquist signaling,'' in \emph{Proc. IEEE
  International Conference on Communications}, May 2017, pp. 1--6.

\bibitem{bedeer2017very}
E.~Bedeer, M.~H. Ahmed, and H.~Yanikomeroglu, ``A very low complexity
  successive symbol-by-symbol sequence estimator for {faster-than-Nyquist}
  signaling,'' \emph{IEEE Access}, vol.~5, pp. 7414--7422, 2017.

\bibitem{kulhandjian2019low}
M.~Kulhandjian, E.~Bedeer, H.~Kulhandjian, C.~D’Amours, and H.~Yanikomeroglu,
  ``Low-complexity detection for faster-than-{N}yquist signaling based on
  probabilistic data association,'' \emph{{IEEE} Commun. Lett.}, vol.~24,
  no.~4, pp. 762--766, Apr. 2020.

\bibitem{ishihara2021evolution}
T.~Ishihara, S.~Sugiura, and L.~Hanzo, ``The evolution of faster-than-{N}yquist
  signaling,'' \emph{IEEE Access}, vol.~9, pp. 86\,535--86\,564, 2021.

\bibitem{whenML}
O.~Simeone, ``A very brief introduction to machine learning with applications
  to communication systems,'' \emph{IEEE Trans. Cogn. Commun. Netw.}, vol.~4,
  no.~4, pp. 648--664, Dec. 2018.

\bibitem{ftnRcv}
P.~Song, F.~Gong, Q.~Li, G.~Li, and H.~Ding, ``Receiver design for
  faster-than-{N}yquist signaling: Deep-learning-based architectures,''
  \emph{IEEE Access}, vol.~8, pp. 68\,866--68\,873, 2020.

\bibitem{ftnSP}
B.~Liu, S.~Li, Y.~Xie, and J.~Yuan, ``A novel sum-product detection algorithm
  for faster-than-{N}yquist signaling: A deep learning approach,'' \emph{IEEE
  Trans. Commun.}, vol.~69, no.~9, pp. 5975--5987, June 2021.

\bibitem{prlja2008receivers}
A.~Prlja, J.~B. Anderson, and F.~Rusek, ``{Receivers for faster-than-Nyquist
  signaling with and without turbo equalization},'' in \emph{Proc. IEEE
  International Symposium on Information Theory}, Jul. 2008, pp. 464--468.

\bibitem{li2017reduced}
S.~Li, B.~Bai, J.~Zhou, P.~Chen, and Z.~Yu, ``{Reduced-complexity equalization
  for faster-than-Nyquist signaling: New methods based on Ungerboeck
  observation model},'' \emph{{IEEE} Trans. Commun.}, vol.~66, no.~3, pp.
  1190--1204, Mar 2018.

\bibitem{6241379}
A.~{Prlja} and J.~B. {Anderson}, ``Reduced-complexity receivers for strongly
  narrowband intersymbol interference introduced by faster-than-{N}yquist
  signaling,'' \emph{IEEE Trans. Commun.}, vol.~60, no.~9, pp. 2591--2601, Sep.
  2012.

\bibitem{textbook}
J.~B. Anderson, \emph{Coded Modulation Systems}.\hskip 1em plus 0.5em minus
  0.4em\relax Kluwer Academic Publishers, 2002.

\bibitem{LLR}
B.~Hochwald and S.~ten Brink, ``Achieving near-capacity on a multiple-antenna
  channel,'' \emph{IEEE Trans. Commun.}, vol.~51, no.~3, pp. 389--399, Apr.
  2003.

\bibitem{mostafa}
M.~Mohammadkarimi, M.~Mehrabi, M.~Ardakani, and Y.~Jing, ``Deep learning-based
  sphere decoding,'' \emph{IEEE Trans. Wirel. Commun.}, vol.~18, no.~9, pp.
  4368--4378, Sep. 2019.

\bibitem{DLBook}
I.~Goodfellow, Y.~Bengio, and A.~Courville, \emph{Deep learning}.\hskip 1em
  plus 0.5em minus 0.4em\relax MIT Press, Nov. 2016.

\bibitem{adam}
D.~P. Kingma and J.~Ba, ``Adam: A method for stochastic optimization,''
  \emph{arXiv preprint arXiv:1412.6980}, Dec. 2014.

\bibitem{hassibi1}
B.~Hassibi and H.~Vikalo, ``On the sphere-decoding algorithm {I}. {E}xpected
  complexity,'' \emph{IEEE Trans. Signal Process}, vol.~53, no.~8, pp.
  2806--2818, July 2005.

\bibitem{ibrahim2021novel}
A.~Ibrahim, E.~Bedeer, and H.~Yanikomeroglu, ``{A novel low complexity
  faster-than-Nyquist signaling detector based on the primal-dual
  predictor-corrector interior point method},'' \emph{{IEEE} Commun. Lett.},
  vol.~25, no.~7, pp. 2370--2374, July 2021.

\end{thebibliography}
\end{document}